\documentclass[11pt]{article}
\usepackage[T1]{fontenc}
\usepackage{lmodern}
\usepackage[margin=1in]{geometry}
\usepackage{amsmath,amssymb,amsthm,mathtools}
\usepackage{booktabs,tabularx,array}
\usepackage{microtype}
\usepackage{needspace}
\usepackage[colorlinks=true,linkcolor=blue!50!black,citecolor=blue!50!black,urlcolor=blue!50!black]{hyperref}
\usepackage{xcolor}
\usepackage{enumitem}
\setlist{nosep}
\newtheorem{theorem}{Theorem}[section]
\newtheorem{lemma}[theorem]{Lemma}
\newtheorem{corollary}[theorem]{Corollary}
\newtheorem{proposition}[theorem]{Proposition}
\theoremstyle{definition}
\newtheorem{definition}[theorem]{Definition}
\newtheorem{assumption}[theorem]{Assumption}
\theoremstyle{remark}
\newtheorem{remark}[theorem]{Remark}
\newcommand{\poly}{\operatorname{poly}}

\newcommand{\OV}{\operatorname{OV}}
\newcommand{\LCS}{\operatorname{LCS}}
\newcommand{\ED}{\operatorname{ED}}
\newcommand{\DTW}{\operatorname{DTW}}
\newcommand{\cP}{\mathcal P}
\DeclareMathOperator{\dist}{dist}
\numberwithin{equation}{section}

\title{Quadratic-Time Barriers from Self-Referential SAT}
\author{Guangyan Zhou\\[0.5em]
\small Department of Mathematics and Statistics, Beijing Technology and Business University, Beijing 100048, China\\
\small \texttt{zhouguangyan@btbu.edu.cn}}
\date{}
\begin{document}
\maketitle
\begin{abstract}
Under the self-referential $K$-SAT exhaustive-search hypothesis (SRK-ESH),
no uniform deterministic algorithm computes exact longest common
subsequence or standard edit distance on binary strings in
$O(n^{2-\eta})$ time for any fixed $\eta>0$, where $n$ is the longer
sequence length. The same exclusion holds for
one-dimensional dynamic time warping on $\{0,1,2,4,8\}$ with
absolute-difference cost and for five-symbol dynamic time warping with
the discrete metric. We derive Orthogonal Vectors hardness in
polylogarithmic dimension by splitting the variables of the
logarithmic-width source formula into two halves. Existing sequence
reductions have polynomial overhead in the dimension. Their output
lengths and construction times allow a fixed saving in the target
exponent to yield a fixed proportional saving in the source exponent.
\end{abstract}

\section{Introduction}
Classical dynamic programs solve longest common subsequence, edit
distance, and dynamic time warping in quadratic time. Fine-grained
reductions connect a fixed improvement in this exponent to faster
algorithms for satisfiability.

The Strong Exponential Time Hypothesis (SETH), originating in the study of satisfiability complexity by Impagliazzo and Paturi~\cite{IP01}, has provided a widely used foundation for these arguments. Backurs and Indyk~\cite{BI18} established the strongly subquadratic barrier for standard edit distance under SETH. Abboud, Backurs, and Vassilevska Williams~\cite{ABW15} and Bringmann and K\"unnemann~\cite{BK15} established reductions connecting vector-pair problems to sequence similarity. These reductions support precise barriers against $O(n^{2-\eta})$ algorithms for a fixed $\eta>0$.

OpenAI's recent preprint~\cite{OpenAI26} gives a deterministic
subexponential-time algorithm for counting satisfying assignments of
$k$-CNF formulas for every fixed $k$, which would refute both ETH and SETH.
Its running-time guarantee is stated for fixed clause width. This raises
the question of whether a hardness hypothesis for logarithmic-width SAT
can support the established quadratic-time barriers for sequence similarity.

The self-referential $K$-SAT exhaustive-search
hypothesis (SRK-ESH) of Fang, Li, Liu, Luo, Tong, and Xu~\cite{FLLLTX26}
excludes fixed proportional exponent savings on a fixed promise family.
We use it to establish the Orthogonal Vectors hardness required by the
reductions behind Table~\ref{tab:retained}. The formal statements also
cover threshold decisions.

\begin{table}[htbp]
\centering\small
\begin{tabularx}{\textwidth}{@{}>{\raggedright\arraybackslash}p{0.18\textwidth}*{2}{>{\raggedright\arraybackslash}X}@{}}
\toprule
Problem & Input restriction & Consequence under SRK-ESH\\
\midrule
Edit distance & Binary strings; unit insertion, deletion, and substitution costs & No deterministic exact $O(n^{2-\eta})$ algorithm\\[3pt]
LCS & Binary strings & No deterministic exact $O(n^{2-\eta})$ algorithm\\[3pt]
One-dimensional DTW & Values in $\{0,1,2,4,8\}$; cost $|a-b|$ & No deterministic exact $O(n^{2-\eta})$ algorithm\\[3pt]
Symbolic DTW & Five symbols; discrete metric & No deterministic exact $O(n^{2-\eta})$ algorithm\\
\bottomrule
\end{tabularx}
\caption{Established quadratic-time barriers retained under SRK-ESH. Each statement holds for every fixed $\eta>0$ and uniform algorithms, for exact values and threshold decisions; $n$ is the maximum sequence length. The first three consequences use the reductions of~\cite{BK15}; the fourth uses~\cite{ABW15}.}
\label{tab:retained}
\end{table}

Splitting the source variables into two halves gives two lists of about
$2^{N/2}$ assignments. Each clause contributes one vector coordinate,
so polynomially many clauses give polylogarithmic dimension in the
list size. Sections~\ref{sec:ov} and~\ref{sec:transfer} account for the
sizes and construction costs along this reduction chain.

\section{SRK-ESH as an alternative hardness assumption}\label{sec:prelim}
\subsection{Parameters and computational conventions}
We use $N$ for the number of variables in the source formula, $M$ for its number of clauses, $q$ for the maximum vector-list size, $d$ for vector dimension, and $n$ for the maximum of the two target sequence lengths. All logarithms with a subscript $2$ are binary. Constants in asymptotic notation may depend on fixed source parameters and on a fixed reduction, but not on the input size.

All algorithms are uniform and deterministic unless stated otherwise. The notation $O^*(\cdot)$ suppresses factors polynomial in the encoded source input length. We interpret the source hypothesis and target bounds in a common random-access model, with standard operations on words of $O(\log L)$ bits for a computation using at most $L$ memory locations. In the hypothetical composed algorithms, word lengths are $O(N)$; charging polynomial bit cost for each such operation only introduces a factor suppressed by $O^*$. No unit-cost operations on unbounded integers are used.

A \emph{strongly subquadratic} algorithm has running time $O(n^{2-\eta})$ for some fixed $\eta>0$. It suffices to exclude $0<\eta<1$, since a faster bound with a larger saving also implies a bound with a smaller positive saving.

\subsection{The self-referential promise family}
Fix the parameters $\varepsilon>0$ and $\lambda>0$ of the logarithmic-width construction in~\cite{FLLLTX26}. Its scale is
\begin{equation}\label{eq:source-scale}
 K_N=\lceil(1+\varepsilon)\log_2 N\rceil,
 \qquad M_N=\Theta(N^{2+\varepsilon}).
\end{equation}
For concreteness, the clause-count tuning uses $p_N=2^{-K_N}$ and
\[
 M_N=\left\lfloor\frac{N\ln 2-\ln\lambda}{-\ln(1-p_N)}\right\rfloor.
\]
Only the polynomial bound on $M_N$ will enter our reductions. A direct literal-list encoding has length $O(M_N K_N\log N)=N^{O(1)}$.

Let $\cP_N$ denote exactly the $N$-variable promise family specified immediately before Assumption~1 in Section~7.2 of~\cite{FLLLTX26}, and put $\cP=\bigcup_N\cP_N$. Its members are the uniquely satisfiable mother formulas and the unsatisfiable self-referential companions described there. A companion is obtained by replacing a redundant clause with a clause, on the same variable support, falsified by the unique satisfying assignment. The formula alone is supplied as input. Neither its satisfying assignment nor a label specifying which member of a SAT/UNSAT pair it represents is supplied. The family is fixed independently of the solver.

We use this family by reference, with its original membership conditions. In particular, no sampling procedure or efficiently recognizable membership test is required for the reductions below: their correctness holds on every CNF formula, and hence on every promised input.

\begin{assumption}[SRK-ESH, Fang et al.~\cite{FLLLTX26}, Assumption~1]\label{ass:srk}
For each fixed $\delta\in(0,1)$, there is no uniform deterministic algorithm $A$ that decides satisfiability correctly on every input in $\cP$ and satisfies
\[
 \max_{F\in\cP_N}T_A(F)=O^*\!\left(2^{(1-\delta)N}\right)
\]
for all sufficiently large $N$, where $T_A(F)$ denotes the running time of $A$ on input $F$.
\end{assumption}

Fang et al.~\cite{FLLLTX26} propose SRK-ESH on the basis of both structural and algorithmic evidence, as reviewed next.

\subsection{Clause width: source hardness and reduction cost}\label{sec:width}
Logarithmic width enters the motivation for SRK-ESH in two ways.
At the threshold in~\eqref{eq:source-scale}, the Poisson limit in
\cite[Theorem~4]{FLLLTX26} gives positive limiting probabilities for
unique satisfiability and unsatisfiability. The redundant-clause
argument and one-clause flip give the self-referential
construction~\cite[Lemma~5 and Theorem~6]{FLLLTX26}. In the fixed-width
ensembles studied in Sections~3--4 of that work, overlap correlations
obstruct the stated solution-independence criterion.
Algorithmic guarantees must also be checked at the growing width
$K_N$. For example, a uniform bound
$f(k)2^{(1-\delta(k))N}N^c$ with fixed $c$ certifies a fixed proportional
saving at $k=K_N$ only if $\delta(K_N)N-\log_2 f(K_N)$ remains
at least a positive constant times $N$, up to polynomial factors.
A width-dependent polynomial exponent or validity threshold also
enters this calculation~\cite[Section~7.2]{FLLLTX26}.

In the reduction, the clause count determines the vector dimension
and the width affects the cost of evaluating each coordinate.
Lemma~\ref{lem:split} gives $q=2^{\lceil N/2\rceil}$,
$d=M_N$, and construction cost $O(qM_N(K_N+1))$ in literal-evaluation
operations. Both $d$ and the multiplicative construction overhead
are polylogarithmic in $q$ under~\eqref{eq:source-scale}.

\subsection{Vector and sequence problems}
\begin{definition}[Orthogonal Vectors]
Given two lists $A,B$ of vectors in $\{0,1\}^d$, each containing at most $q$ indexed vectors, $\OV(A,B)$ asks whether there are $a\in A$ and $b\in B$ with
\[
 a\cdot b=\sum_{j=1}^d a_jb_j=0.
\]
Repeated vectors are allowed and are distinguished by their list indices.
\end{definition}

For strings $x,y$, $\LCS(x,y)$ is the maximum length of a sequence that is a subsequence of both. The standard edit distance $\ED(x,y)$ is the minimum number of single-symbol insertions, deletions, and substitutions needed to transform $x$ into $y$, each with unit cost. Our LCS and edit-distance results use the binary alphabet.

A DTW traversal of two nonempty sequences is a path from $(1,1)$ to $(|x|,|y|)$ using steps $(1,0)$, $(0,1)$, and $(1,1)$. Its cost is the sum of $\dist(x_i,y_j)$ over visited positions, and $\DTW(x,y)$ is the minimum traversal cost. For the main DTW result, entries lie in $\{0,1,2,4,8\}\subset\mathbb R$ and $\dist(a,b)=|a-b|$. A separate corollary uses a five-symbol alphabet with discrete metric $\dist(a,b)=\mathbf{1}[a\ne b]$. These are different metric spaces. The target size is $n=\max\{|x|,|y|\}$. We consider exact values or threshold decisions: $\LCS(x,y)\ge t$, $\ED(x,y)\le t$, and $\DTW(x,y)\le t$.

\section{Establishing the Orthogonal Vectors barrier}\label{sec:ov}
We establish the OV hardness required by the sequence reductions, applying split and list directly to the source formulas.

\subsection{The split-and-list construction}
The following standard construction is stated for arbitrary CNF formulas so that its application to the source promise is explicit.

\begin{lemma}[Split and list]\label{lem:split}
Let $F=C_1\wedge\cdots\wedge C_M$ be a CNF formula on $N$ variables, with clauses of width at most $K$. One can construct two lists of $M$-dimensional Boolean vectors, of sizes $2^{\lfloor N/2\rfloor}$ and $2^{\lceil N/2\rceil}$, such that $F$ is satisfiable if and only if the lists contain an orthogonal pair. The construction takes
\[
 O\!\left(2^{\lceil N/2\rceil}M(K+1)\right)
\]
elementary literal-evaluation operations, up to polynomial encoding overhead. Before any list padding, the number of indexed orthogonal pairs equals the number of satisfying assignments of $F$.
\end{lemma}
\begin{proof}
Partition the variables into $X$ and $Y$ of sizes $\lfloor N/2\rfloor$ and $\lceil N/2\rceil$. For each assignment $\alpha$ to $X$, define $a_\alpha\in\{0,1\}^M$ by
\[
 a_\alpha[j]=
 \begin{cases}
 0,&\text{some literal of $C_j$ on a variable in $X$ is true under $\alpha$},\\
 1,&\text{otherwise}.
 \end{cases}
\]
In particular, if $C_j$ contains no variable from $X$, then $a_\alpha[j]=1$ for every $\alpha$. Define $b_\beta$ analogously for assignments $\beta$ to $Y$. A clause containing no variable from one half has coordinate $1$ for every assignment to that half.

For every pair $(\alpha,\beta)$ and every clause index $j$, the product $a_\alpha[j]b_\beta[j]$ is $1$ exactly when neither half makes a literal of $C_j$ true. Consequently,
\begin{equation}\label{eq:unsat-count}
 a_\alpha\cdot b_\beta
 =\bigl|\{j:(\alpha,\beta)\text{ falsifies }C_j\}\bigr|.
\end{equation}
The inner product vanishes exactly when the combined assignment satisfies $F$. Full assignments correspond bijectively to indexed pairs of half-assignments, proving also the counting statement.

For each half-assignment, scan all clauses and at most $K$ literals per clause. There are at most $2q$ half-assignments, where $q=2^{\lceil N/2\rceil}$. This gives the claimed construction bound. The lists occupy $O(qM)$ bits, apart from indexing overhead.
\end{proof}

If a downstream construction requires equal list sizes, repeat an existing vector in the smaller list until both sizes equal $q$. This preserves the existence of an orthogonal pair. The counting assertion of Lemma~\ref{lem:split} concerns the original, unpadded lists.

\paragraph{A small example.}
Consider
\[
 F=(x_1\vee y_1)\wedge(\neg x_1\vee y_2)\wedge(x_2\vee\neg y_1).
\]
For $\alpha=(x_1,x_2)=(0,1)$, the vector is $a_\alpha=(1,0,0)$. For $\beta=(y_1,y_2)=(1,0)$, it is $b_\beta=(0,1,1)$. Their inner product is zero and the combined assignment satisfies $F$. If instead $\beta=(0,0)$, then $b_\beta=(1,1,0)$ and the inner product is one, corresponding to the falsified first clause. This example illustrates the encoding; it is not intended as a member of the source promise family.

\subsection{Dimension and conditional hardness}
\begin{theorem}[OV barrier from SRK-ESH]\label{thm:ov}
Fix the source parameters in Assumption~\ref{ass:srk}. There is a constant $C>0$ such that, under SRK-ESH, no uniform deterministic algorithm solves all OV instances with
\begin{equation}\label{eq:dim}
 d\le C(1+\log_2 q)^{2+\varepsilon}
\end{equation}
in time $O(q^{2-\eta}d^c)$ for any fixed $\eta>0$ and $c\ge0$.
\end{theorem}
\begin{proof}
Apply Lemma~\ref{lem:split} to a promised formula $F\in\cP_N$. The resulting instance has
\begin{equation}\label{eq:parameters}
 q=2^{\lceil N/2\rceil},\qquad d=M_N=\Theta(N^{2+\varepsilon}).
\end{equation}
Because $N\le2\log_2q$, a fixed $C$ makes~\eqref{eq:dim} valid for all sufficiently large $N$. Enlarge $C$ for finitely many exceptions if needed.

Suppose the claimed OV algorithm exists. Reduce the saving if necessary so that $0<\eta<1$. The construction followed by that algorithm decides every promised formula in time
\begin{align*}
 O^*(2^{N/2})+O(q^{2-\eta}d^c)
 &=O^*(2^{N/2})+O^*\!\left(2^{(1-\eta/2)N}\right)\\
 &=O^*\!\left(2^{(1-\eta/2)N}\right).
\end{align*}
The rounding in $q$ changes only a constant factor. The constructed algorithm is uniform, deterministic, and correct on the whole promise family. Taking $\delta=\eta/2$ contradicts Assumption~\ref{ass:srk}. Finitely many small variable counts can be handled by exhaustive search.
\end{proof}

Theorem~\ref{thm:ov} specifies a dimension regime, rather than identifying all formulations of an OV hypothesis. In particular, it does not by itself establish the corresponding claim in dimension $O(\log q)$. Its polylogarithmic dimension is exactly what the applications below require.

\begin{remark}[The inherited promise]
On the unpadded images of $\cP_N$, there is either one indexed orthogonal pair or none. This follows from Lemma~\ref{lem:split} and the SAT/UNSAT promise. The source contradiction already applies to an algorithm correct on those images alone. Our target lower bounds concern algorithms correct on all target inputs, which necessarily include these images. No efficient recognition or sampling of the image family is used.
\end{remark}

\section{Retaining quadratic-time barriers}\label{sec:transfer}
We now verify that the existing sequence reductions meet a transfer criterion that charges both construction and solution time.

\subsection{Transfer with polynomial dimension overhead}
\begin{theorem}[Quadratic transfer]\label{thm:transfer}
Assume SRK-ESH. Let $n$ denote the size parameter of a target problem $Q$. Suppose a uniform deterministic reduction maps two lists of $q$ Boolean vectors in dimension $d$ to one $Q$ instance, with output size and construction time bounded by $q\,d^{O(1)}$. Suppose also that the OV answer is recoverable from the exact $Q$ answer within $q\,d^{O(1)}$ additional time. Then $Q$ admits no uniform deterministic $O(n^{2-\eta})$ algorithm for any fixed $\eta>0$.
\end{theorem}
\begin{proof}
Reduce $F\in\cP_N$ to OV using Lemma~\ref{lem:split}. Then $q=2^{\lceil N/2\rceil}$ and $d=N^{O(1)}$. Apply the reduction to $Q$ and run the hypothetical solver. For $0<\eta<1$, the total time, including construction and decoding, is
\[
 O^*(2^{N/2})+O^*\!\left((2^{N/2})^{2-\eta}\right)
 =O^*\!\left(2^{(1-\eta/2)N}\right).
\]
This contradicts Assumption~\ref{ass:srk}. Larger positive savings imply a bound with a saving in $(0,1)$.
\end{proof}

The following extension gives a criterion for testing further target reductions with larger size expansion.
\begin{proposition}[General expansion]\label{prop:expansion}
Assume SRK-ESH. Suppose the reduction and decoding in Theorem~\ref{thm:transfer} instead have size and time bounds $q^\gamma d^{O(1)}$, where $1\le\gamma<2$ is fixed. Then no uniform deterministic $O(n^\rho)$ algorithm exists for $Q$ when $\rho>0$ and $\gamma\rho<2$.
\end{proposition}
\begin{proof}
The complete source running time is
\[
 O^*\!\left(2^{N/2}+2^{\gamma N/2}+2^{\gamma\rho N/2}\right).
\]
With $\theta=\max\{1/2,\gamma/2,\gamma\rho/2\}<1$, this is $O^*(2^{\theta N})$, contradicting SRK-ESH. An output-size bound without the construction-time bound would not justify this argument.
\end{proof}

\subsection{The sequence reduction interface}
\begin{lemma}[Bringmann--K\"unnemann~\cite{BK15}]\label{lem:bk}
OV on two lists of $q$ vectors in $\{0,1\}^d$ reduces to binary LCS, standard edit distance on binary strings, and one-dimensional DTW over $\{0,1,2,4,8\}$ with local cost $|a-b|$. Each reduction has output length and construction time $O(qd)$; an output threshold decides OV.
\end{lemma}

\paragraph{Dimension accounting.}
The proof of Theorem~III.3 in~\cite{BK15}, Section~III-A, constructs coordinate, vector, and normalized vector gadgets and then the final sequences. Its three alignment-gadget levels have total cost $O((q_1+q_2)d)$ for list sizes $q_1,q_2$. Lemmas~V.2 and~V.4 provide binary edit-distance gadgets with insertion and deletion costs $1$, matching cost $0$, and any fixed substitution cost in $(0,2]$. Substitution cost $1$ gives standard edit distance; cost $2$ gives the LCS interface. Lemmas~IV.1 and~IV.3 provide the DTW interface. The latter's three levels introduce values $2,4,8$, starting from $0,1$. Thus both symbol encoding and the number of levels are independent of $d$.

The bound $O(qd)$ applies at $d=M_N$. The LCS interface minimizes
$|x|+|y|-2\LCS(x,y)$, which converts its decision threshold into
an LCS threshold.

\subsection{Conditional lower bounds}
The edit-distance conclusion retains the barrier established by Backurs and Indyk~\cite{BI18}, using the binary-alphabet refinement of Bringmann and K\"unnemann~\cite{BK15} as the reduction interface.

\begin{theorem}[Sequence similarity under SRK-ESH]\label{thm:main}
Assume SRK-ESH as introduced in~\cite{FLLLTX26}. For every fixed $\eta>0$, none of the following admits a uniform deterministic exact algorithm with running time $O(n^{2-\eta})$:
\begin{enumerate}[label=(\roman*)]
\item standard edit distance of two binary strings, with unit insertion, deletion, and substitution costs;
\item LCS of two binary strings;
\item DTW of two one-dimensional sequences with values in $\{0,1,2,4,8\}$ and local cost $|a-b|$.
\end{enumerate}
Here $n$ is the maximum of the two sequence lengths. The same exclusion holds for their threshold decision versions with the threshold supplied as input.
\end{theorem}
\begin{proof}
Lemma~\ref{lem:bk} supplies the required $O(qd)$ reduction for each problem. Apply Theorem~\ref{thm:transfer}. For threshold decision, use the threshold supplied by the reduction, with the appropriate comparison direction.
\end{proof}

\begin{corollary}[Discrete-metric DTW]\label{cor:discrete}
Under SRK-ESH, exact DTW over five symbols with the discrete metric, and its threshold decision version, admit no uniform deterministic $O(n^{2-\eta})$ algorithm for any fixed $\eta>0$.
\end{corollary}
\begin{proof}
The OV-to-DTW construction of Abboud, Backurs, and Vassilevska Williams~\cite{ABW15}, Section~IV and Theorem~7, has size and construction cost $q\,d^{O(1)}$ and supplies a decision threshold. Apply Theorem~\ref{thm:transfer}.
\end{proof}

The complete parameter calculation for $0<\eta<1$ is
\begin{equation}\label{eq:complete}
\begin{gathered}
 F\in\cP_N
 \ \longmapsto\ (q,d)=\left(2^{\lceil N/2\rceil},N^{O(1)}\right)
 \ \longmapsto\ n\le 2^{N/2}N^{O(1)},\\
 O(n^{2-\eta})\ \longmapsto\ O^*\!\left(2^{(1-\eta/2)N}\right).
\end{gathered}
\end{equation}
Polynomial factors in $d$ remain polynomial in the original variable count $N$.

\begin{table}[t]
\centering
\small
\begin{tabularx}{\textwidth}{@{}lXX@{}}
\toprule
Stage & Size parameters & Time or consequence\\
\midrule
Source promise & $N$ variables; $M_N=\Theta(N^{2+\varepsilon})$ & SRK-ESH excludes $O^*(2^{(1-\delta)N})$\\[3pt]
Split and list & $q=2^{\lceil N/2\rceil}$; $d=M_N$ & Construct in $2^{N/2}\poly(N)$\\[3pt]
Main sequence reductions & $n=O(qd)$ & Construct in $O(qd)$\\[3pt]
Hypothetical solver & $O(n^{2-\eta})$ & Source time $O^*(2^{(1-\eta/2)N})$\\
\bottomrule
\end{tabularx}
\caption{Parameter accounting for Theorem~\ref{thm:main}. All savings are fixed constants; polynomial factors in $N$ are suppressed only at the source exponential scale.}
\label{tab:parameters}
\end{table}

\section{Scope of the bounds}\label{sec:discussion}
The bounds apply to exact values and threshold decisions for the input
versions in Table~\ref{tab:retained}. They exclude a fixed power saving
in the worst-case running time of a uniform deterministic algorithm.
For example, equation~\eqref{eq:complete} converts an $O(n^{1.9})$
algorithm into an $O^*(2^{0.95N})$ source algorithm. An improvement
to $n^2/\log^c n$ for fixed $c$ falls outside this exclusion.

\section{Conclusion}
SRK-ESH yields the deterministic strongly subquadratic barriers for
binary LCS, binary standard edit distance, and the two stated DTW
variants. The split-and-list construction supplies Orthogonal Vectors
instances in the dimension regime used by the sequence reductions.

The same transfer criterion applies to other targets. The required
vector dimension, output size, construction time, and answer recovery
must leave a fixed proportional saving in the source exponent when
combined with the proposed target algorithm.


\begingroup
\small
\begin{thebibliography}{9}
\setlength{\itemsep}{3pt}
\bibitem{ABW15}
Amir Abboud, Arturs Backurs, and Virginia Vassilevska Williams.
\newblock \emph{Tight Hardness Results for LCS and Other Sequence Similarity Measures}.
\newblock In \emph{Proceedings of the 56th Annual IEEE Symposium on Foundations of Computer Science (FOCS)}, pages 59--78. IEEE, 2015.
\newblock \url{https://doi.org/10.1109/FOCS.2015.14}.

\bibitem{BK15}
Karl Bringmann and Marvin K\"unnemann.
\newblock \emph{Quadratic Conditional Lower Bounds for String Problems and Dynamic Time Warping}.
\newblock In \emph{Proceedings of the 56th Annual IEEE Symposium on Foundations of Computer Science (FOCS)}, pages 79--97. IEEE, 2015.
\newblock \url{https://doi.org/10.1109/FOCS.2015.15}.

\bibitem{BI18}
Arturs Backurs and Piotr Indyk.
\newblock Edit Distance Cannot Be Computed in Strongly Subquadratic Time (Unless SETH Is False).
\newblock \emph{SIAM Journal on Computing}, 47(3):1087--1097, 2018.
\newblock \url{https://doi.org/10.1137/15M1053128}.

\bibitem{FLLLTX26}
Wen Fang, Xianxian Li, Jun Liu, Jie Luo, Yongxin Tong, and Ke Xu.
\newblock \emph{Self-Referential $K$-SAT and the Finite Analogue of G\"odel's Incompleteness Theorem}.
\newblock arXiv:2607.01671v3, October 29, 2026.
\newblock Section~7.2, Assumption~1 (SRK-ESH).
\newblock \url{https://arxiv.org/abs/2607.01671v3}.

\bibitem{IP01}
Russell Impagliazzo and Ramamohan Paturi.
\newblock On the complexity of $k$-SAT.
\newblock \emph{Journal of Computer and System Sciences}, 62(2):367--375, 2001.
\newblock \href{https://doi.org/10.1006/jcss.2000.1727}{doi:10.1006/jcss.2000.1727}.
\bibitem{OpenAI26}
OpenAI.
\newblock The Exponential Time Hypothesis is False.
\newblock Preprint, revised October 25, 2026.
\newblock \href{https://github.com/openai/math/blob/main/preprints/The-Exponential-Time-Hypothesis-is-False-October-25-2026/paper.pdf}{OpenAI Math Release}.
\end{thebibliography}
\endgroup
\end{document}